\documentclass{optica-article}

\journal{ome}

\articletype{Research Article}
\usepackage{color}

\usepackage{lineno}

\begin{document}

\title{Localization landscape of optical waves in multifractal photonic membranes}

\author{Tornike Shubitidze\authormark{1,$\dagger$}, Yilin Zhu\authormark{2,$\dagger$}, Hari Sundar\authormark{4}, and Luca Dal Negro\authormark{1-3,*}}

\address{\authormark{1}Department of Electrical \& Computer Engineering and Photonics Center, Boston University, 8 Saint Mary's Street, Boston, Massachusetts 02215, USA\\
\authormark{2}Division of Materials Science \& Engineering, Boston University, 15 Saint Mary's Street, Boston, Massachusetts, 02215, USA\\
\authormark{3}Department of Physics, Boston University, 590 Commonwealth Avenue, Boston, Massachusetts 02215, USA\\
\authormark{4}Kahlert School of Computing, University of Utah, 50 Central Campus Drive, Salt Lake City, Utah 84112, USA\\
\authormark{$\dagger$}These authors contributed equally to the work. }

\email{\authormark{*}dalnegro@bu.edu} 



\begin{abstract}
In this paper, we investigate the localization properties of optical waves in disordered systems with multifractal scattering potentials. In particular, we apply the localization landscape theory to the classical Helmholtz operator and, without solving the associated eigenproblem, show accurate predictions of localized eigenmodes for one- and two-dimensional multifractal structures. Finally, we design and fabricate nanoperforated photonic membranes in silicon nitride (SiN) and image directly their multifractal modes using leaky-mode spectroscopy in the visible spectral range. The measured data demonstrate optical resonances with multiscale intensity fluctuations in good qualitative agreement with numerical simulations. The proposed approach provides a convenient strategy to design multifractal photonic membranes, enabling rapid  exploration of extended scattering structures with tailored disorder for enhanced light-matter interactions.
\end{abstract}

\section{Introduction}
In the last two decades, the search for Anderson light localization \cite{anderson1958absence} and the study of disorder-induced phenomena for optical waves has stimulated the growing field of ``disordered photonics'', resulting in a wide range of applications to light generation and random lasing \cite{wiersma2008physics}, solar energy \cite{vynck2012photon}, imaging and spectroscopy \cite{mosk2012controlling,redding2013compact}, nonlinear and quantum photonics \cite{dal2022waves,wiersma2013disordered,sapienza2010cavity}. 

Structurally complex photonic structures with non-periodic refractive index variations on the wavelength scale display a very rich physics driven by wave interference effects in the multiple scattering regime with profound analogies to the transport of electrons in disordered metallic alloys and semiconductors \cite{lagendijk2009fifty,akkermans2007mesoscopic}. In particular, various mesoscopic phenomena known for the electron transport in disordered materials, such as the weak localization of light \cite{wolf1985weak}, universal conductance fluctuations \cite{scheffold1998universal}, and Anderson  localization have found their counterparts in disordered optical materials as well \cite{lagendijk2009fifty,anderson1958absence}. 
As originally understood by Philip Anderson in the context of metal-insulator transitions \cite{anderson1958absence}, the transport of quantum waves in strongly disordered media can be completely inhibited by exponentially localized eigenmodes of the Schr\"odinger equation \cite{thouless1974electrons}. Moreover, even much weaker disorder can  significantly modify the traditional Boltzmann transport picture due to recurrent scattering events that already occur in the so-called weak localization regime \cite{wolf1985weak}. Under these conditions, the diffusion constant is modified (i.e., renormalized) by interference phenomena giving rise to weakly-localized eigenmodes with a slow amplitude decay and large intensity fluctuations \cite{sgrignuoli2020subdiffusive}. 

Recently, considerable progress has been made to understand the universal mechanisms of both weak and strong (Anderson) localization and efficiently predict localized modes in disordered structures based on the localization landscape theory \cite{Filoche2012PNAS,Lyra2015EDP,Arnold2019SIAM}. This general mathematical approach solves a Dirichlet problem defined by the uniform forcing of the system Hamiltonian and has been successfully utilized to accurately pinpoint the spatial
locations of localized modes for any positive-definite disordered potential. Introducing the concept of an effective localization potential \cite{Arnold2019SIAM}, researchers have been able to deterministically identify and characterize the eigenvalues and the subregions of a random medium that support localized eigenfunctions without having to solve computationally prohibitive eigenproblems. However, these current landscape-based methods are mostly focused on studying the localization of quantum waves that solve the Schr\"{o}dinger equation and the question naturally arises whether the mathematical landscape theory can be extended to predict the general behavior of classical waves in complex nanophotonic structures. 

Recently, Anderson localization has been explored using the landscape theory in the context of classical acoustic waves by performing a Webster transformation that converts the classical wave equation into an effective Schr\"{o}dinger equation with the same localization properties \cite{colas2022crossover}. These findings suggest a path to exploit the landscape analysis in the study of optical waves governed by the scalar Helmholtz operator, potentially enabling the rapid design and prototyping of extended photonic structures with arbitrary random potentials. In particular, this approach responds to the growing need to investigate the general localization behavior of complex optical potentials with tailored disorder beyond uncorrelated  randomness. 

In order to establish the potential of the localization landscape approach for the design of large-scale photonic structures with correlated disorder, we address here the optical resonances of multi-particle scattering systems with multifractal geometry. These types of random media are described by a continuous distribution of local scaling exponents that characterizes their distinctive fluctuation properties and have recently attracted a growing interest in various scientific fields ranging from finance \cite{calvet2002multifractality}, optical scattering \cite{Chen2023PRB} and even contemporary arts \cite{mureika2005multifractal}. They add novel functionalities to the manipulation of optical fields in complex media \cite{Chen2023PRB} beyond periodic \cite{Joannopoulos2008photonic} or uncorrelated disordered systems \cite{wiersma2013disordered,lagendijk2009fifty}, with emerging applications to active nano-devices and metamaterials \cite{aslan2016multispectral}.

In this article, we propose a general method to calculate the landscape function of classical optical waves in multifractal structures based on the Helmholtz equation. First, we formulate and solve an eigenproblem analogous to the Hamiltonian operator of the Schr\"{o}dinger equation and validate our results by considering the well-known problems of one-dimensional waveguide structures. We then apply the developed approach to both one-dimensional (1D) and two-dimensional (2D) multifractal systems. Next, we discuss the generation of the investigated multifractal scattering structures and then we calculate the associated localization landscapes by solving the Dirichlet problem for the classical optical potential. Based on the obtained Helmholtz localization landscapes, we design and fabricate nanoperforated multifractal membranes in SiN and investigate experimentally their optical resonances using leaky-mode imaging spectroscopy. 

\section{Generation of multifractal scattering structures}

In order to compute the landscape function and the eigenmodes of 1D multifractal potentials we considered two canonical models of multifractal behavior: (i) the binominal multiplicative cascade, which is based on  multiplicative random processes \cite{anitas2019small,calvet2002multifractality} and (ii) the multifractal model of asset returns (MMAR), originally developed for the analysis of financial markets \cite{calvet2002multifractality,mandelbrot1997multifractal}. 

A random multiplicative cascade model is constructed based on a probability field obtained by first dividing an interval into two equal subintervals. Inside each subinterval, one assigns the probabilities $p_{i}\in[0, 1]$ with $i = 1, 2$. This constitutes the first iteration of the process ($n=1$). At the iteration step $n = 2$, each of the two previous subintervals is further divided in two smaller ones, and the probabilities associated with each sub-division are multiplied in random order (i.e., after random reshuffling) with the ones of the previous iterations.

At the iteration $n=3$, one performs a similar division into subintervals and to each of them assigns the probabilities in random permutations from the previous iteration steps. This recurrent  multiplicative cascade distribution defines a multifractal probability field in the limit of a large number of iterations \cite{martinez1990clustering}.
The probability value attached to a square region is the product of the $p_{i}$'s of the square and of all its ancestors at previous generations. The distribution of cell values strictly depends on the initial choice of the probability vector $p$ that acts as a control parameter for the resulting structures. 
The corresponding multiplicative random process is generally non-Gaussian \cite{Chen2023PRB}. 
One-dimensional point patterns (i.e., point processes) with multifractal scaling properties are induced by the probability fields described above by distributing $N$ particles on the specified line segment with probabilities that are proportional to the subinterval values. We achieved this goal using the Monte Carlo rejection scheme and we generated multifractal arrays with $N=1024$ point scatterers positioned along a line segment of length $L=1024\,\mu\mathrm{m}$. To generate the arrays, we chose the initial probability array $p=[0.4,0.6]$. 

The robustness of the proposed landscape method for the Helmholtz operator is supported by additionally considering multifractal point patterns generated based on the MMAR model \cite{calvet2002multifractality,mandelbrot1997multifractal}. The MMAR is a continuous-time process $X(t)$ that captures the heavy tails and long-memory volatility persistence of asserts often exhibited by realistic  financial data \cite{calvet2002multifractality}. It is constructed by compounding a Brownian motion $B(t)$ with a random increasing function $\theta(t)$ according to: 
\begin{equation}\label{eq:logprice}
	X(t)\equiv\ln{P(t)}-\ln{P(0)}=B[\theta(t)]
\end{equation}
where $P(t)$ is the price of a financial asset at a specific time and $\theta(t)$, known as the trading time function, is the cumulative distribution function (CDF) of a multifractal measure $\mu$. In our example we compound with respect to the binomial multiplicative cascade introduced before.

We computed the MMAR over a time series of length $T=1024$, and assigned point scatterers along the time series using the Monte Carlo rejection method, obtaining a multifractal optical potential of a length $N=1024$. Readers can find additional details on the MMAR methods in the specialized literature \cite{calvet2002multifractality,mandelbrot1997multifractal}. 

\section{The Helmholtz localization landscape of one-dimensional multifractals}

We now introduce the localization landscape approach for the Helmholtz operator in typical  one-dimensional multifractal (1D) systems. The approach discussed here for random multifractal potentials can also be applied to deterministic scattering potentials with aperiodic order, which have been extensively investigated in the literature for the engineering of complex optical devices compatible with standard nanofabrication technology \cite{macia2012exploiting,dal2022waves,dal2022waveloc}.

Our goal is to apply the localization landscape theory (LLT) to photonic problems governed by the scalar
Helmholtz equation in a scattering dielectric medium with arbitrary geometry. We start from the Helmholtz equation in a non-homogeneous medium:
\begin{equation}\label{eq:1DHelmholtz}
    \left[\Delta + k_0^2\epsilon(x) \right]\psi=0
\end{equation}
where $k_0$ is the free-space wavenumber and $\epsilon(x)$ is the spatially varying permittivity of the medium. We then
define the variation from the background medium in terms of the optical scattering potential $ V_s(x)$ as follows \cite{dal2022waves}:
\begin{equation}
    k^2=k_0^2\epsilon(x)=k_b^2-V_s(x)
\end{equation}
where $k^2_b=k_0^2\epsilon_b$, and $\epsilon_b$ is the permittivity of the background medium (we use $\epsilon_b=1$). From the expression above we have:
\begin{equation}\label{eq:1Dopt_sca_potential}
    V_s(x)=k_0^2\left[\epsilon_b-\epsilon(x)\right]. 
\end{equation}
We can now formally express the Helmholtz equation as the equivalent Schr\"{o}dinger equation:
\begin{equation}\label{eq:1Deigenproblem}
    \left[-\Delta + V_s(x) \right]\psi=k_0^2\epsilon_b\psi.
\end{equation}

However, the potential  $V_s(x)$ can become negative when the permittivity of the scattering
particles is larger than the one of the background (host medium). Therefore, in order to apply the LLT we
implement the following transformation:
\begin{equation}
    V_s(x)\rightarrow V'(x)\equiv V_s(x)+k_0^2\left[\epsilon_\mathrm{max}-\epsilon_b\right]=k_0^2\left[\epsilon_\mathrm{max}-\epsilon(x)\right].
\end{equation}
were $\epsilon_\mathrm{max}$ is the maximum value of the permittivity of the scattering medium. The shift introduced above renders the potential non-negative, which enables the application of the LLT to general optical wave problems without affecting their physical solutions.  
We now introduce the
Helmholtz landscape equation based on the Dirichlet solution of the equation:
\begin{equation}\label{eq:1Dlandscape}
    Hu(x)=1
\end{equation}
where we defined the optical Hamiltonian $H=-\Delta + V'(x)$. Equation \ref{eq:1Dlandscape} allows us to investigate the localization landscape function $u(x)$ that encodes fundamental information on the localized eigenmodes  \cite{Filoche2012PNAS,Lyra2015EDP,Arnold2019SIAM}.

\begin{figure}[h]
\includegraphics[width=1.0\linewidth]{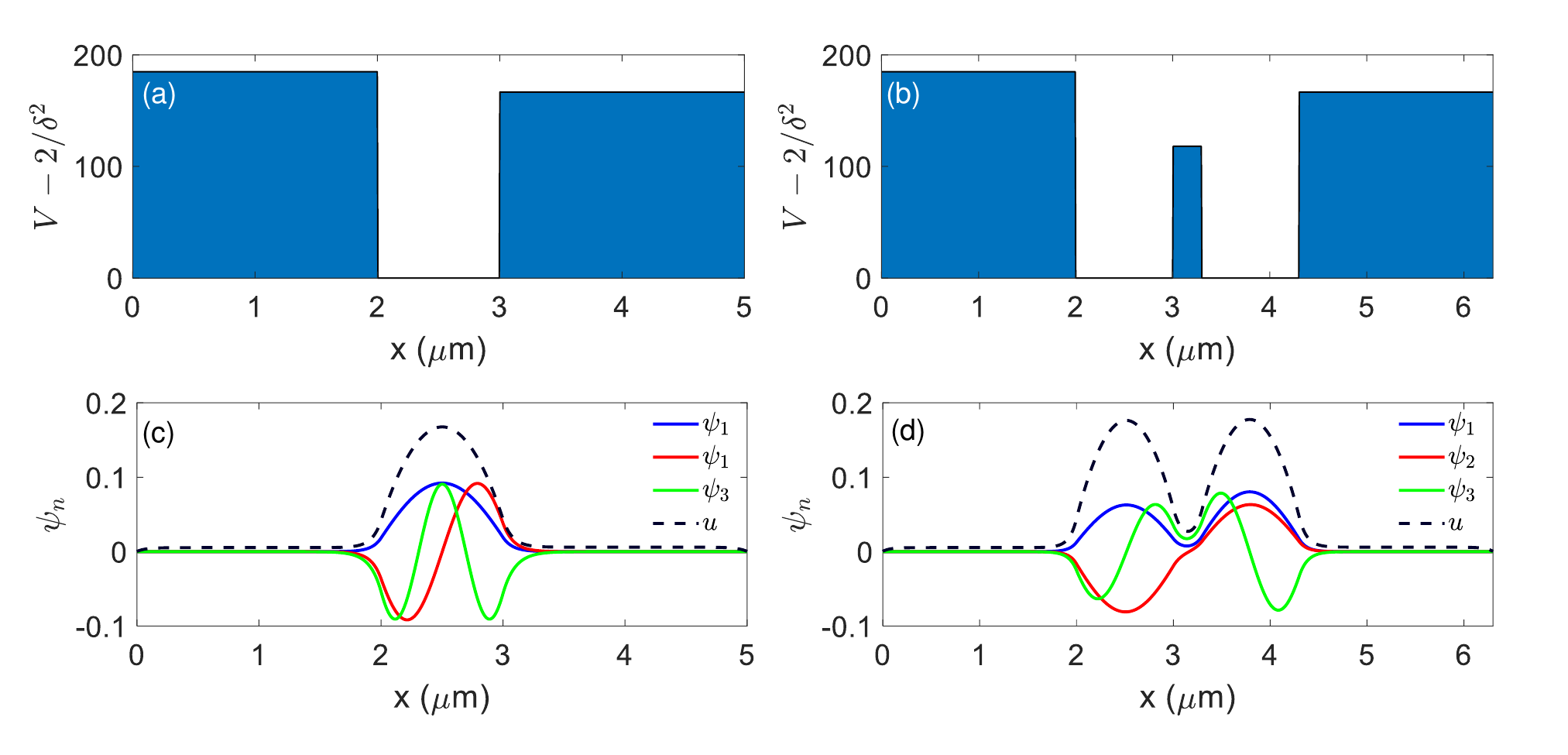}
\centering\caption{\label{fig:waveguide} Tight-binding optical potential of (a) asymmetric slab waveguide and (b) coupled slab waveguides system. (c,d) Spatial profiles of the lowest three optical modes from the solution of Eq. \ref{eq:1Deigenproblem} along with the calculated landscape function $u$ (doted lines) computed from Eq. \ref{eq:1Dlandscape}. }
\end{figure}

In order to numerically calculate the landscape function, we discretize the 1D potential function by considering the
tight-binding problem of a single particle restricted to move along a discrete chain with first-neighbor hopping rate $t$, on-site optical scattering potentials $V_{i}$, and we denote by $u_i$ the landscape function value at site $i$. Using the central difference formula \cite{Lyra2015EDP}, we can approximate Eq. \ref{eq:1Dlandscape} in the following form:
\begin{equation}\label{eq:1Dcentral_difference}
    -t\frac{\left[u_{i+1} + u_{i-1} - 2u_i\right]}{\delta^2}+ \left(V_i - \frac{2t}{\delta^2}\right)u_i=1
\end{equation}
where $\delta$ is the discretization step of the coordinate $x$-axis. To ensure that the spectrum of the possible eigenenergies is positive everywhere, the optical scattering potential term should satisfy the additional condition that $V_i-2t/\delta^2\geq0$ \cite{Lyra2015EDP}. Note that since the potential $V^{\prime}$ is a bounded non-negative function, one can also define the so-called "effective potential" $W=1/u$. The profiles of the effective
potential $W$ are often more regular than the optical potential and display clear structures of walls and wells that serve to identify the regions of localization of the eigenmodes. A rigorous connection has been made between the exponential decay of eigenfunctions and the effective potential $W$ \cite{Arnold2019SIAM}:
\begin{equation}\label{eq:eigfunc_landscape}
    \int_\Omega{e^{h(x)}\psi^2 \mathrm{d}x}\leq C\int_\Omega{\psi^2 \mathrm{d}x}
\end{equation}
where $C$ is a constant, and $h(x)$ is defined as the Agmon's distance \cite{agmon2014lectures} from $x$ to a subset $S$ of $W$ defined as $S=\left\{x\in\Omega\,|\, W(x)\leq\lambda+\delta\right\}$ for any $\delta>0$, where $\lambda$ is the eigenvalue. This enables the accurate prediction of the geometrical supports of the localized eigenfunctions, which are centered at the locations of "wells" (i.e., local minima) of $W$ and are bounded by the heights of the "walls" proportional to the eigenvalues \cite{Arnold2019SIAM}.
\begin{figure}[h]
	\includegraphics[width=0.9\linewidth]{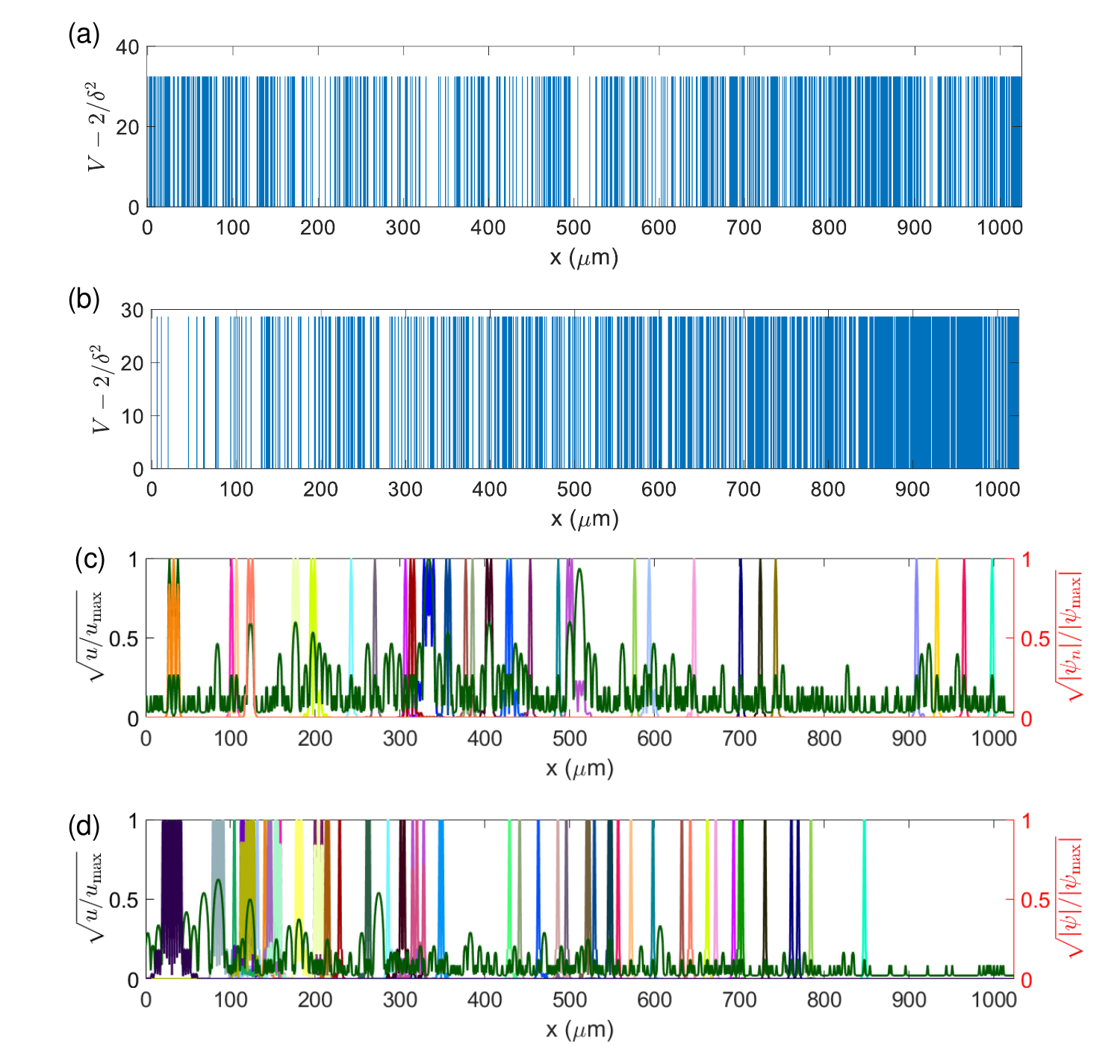}
	\centering\caption{\label{fig:1Dcascade} Constructed multifractal potential by (a) the 1D multiplicative cascade method and (b)the MMAR method. Also shown in (c) and (d) are square root of landscape functions, normalized to their maxima corresponding to (a) and (b), calculated for $\overline{d}_1/\lambda=0.4$. First five modes of each potential are overlayed on top of the landscape function, normalized to their intensities respectively. } 
\end{figure}

In order to demonstrate the predictive power of the Helmholtz landscape introduced above, we consider first the simple case of a three-layer asymmetric waveguide structure with a dielectric permittivity in the core region $\epsilon_{c}=\epsilon_\mathrm{max}=12.25$ and with cladding permittivities $\epsilon_{1}=1.00$ and $\epsilon_{2}=2.10$. We use a wavelength $\lambda=1.55\,\mu$m and display the tight-binding optical potential from Eq. \ref{eq:1Dopt_sca_potential} and Eq. \ref{eq:1Dcentral_difference} in Figure \ref{fig:waveguide}(a). In addition, we considered a system of two coupled dielectric waveguides with the optical potential shown in Figure \ref{fig:waveguide}(b). These are canonical eigenproblems in photonics whose solutions can easily be obtained with traditional methods \cite{wartak2013computational}. Here, by computing the eigensolutions of the corresponding Schr\"{o}dinger Eq. \ref{eq:1Deigenproblem}, we plot the first three optical modes of the two investigated systems in Figures \ref{fig:waveguide}(c,d), respectively. The dashed lines indicate the corresponding Helmholtz landscapes obtained by solving Eq. \ref{eq:1Dlandscape}. In all the calculations we set the hopping rate to be $t=1$, without loss of generality. The results plotted in Figure \ref{fig:waveguide} indicate that the Helmholtz landscape accurately predicts the localization regions of the optical modes computed from the tight-binding solution of the equivalent Schr\"{o}dinger problem. 
\begin{figure}[h]
	\includegraphics[width=0.9\linewidth]{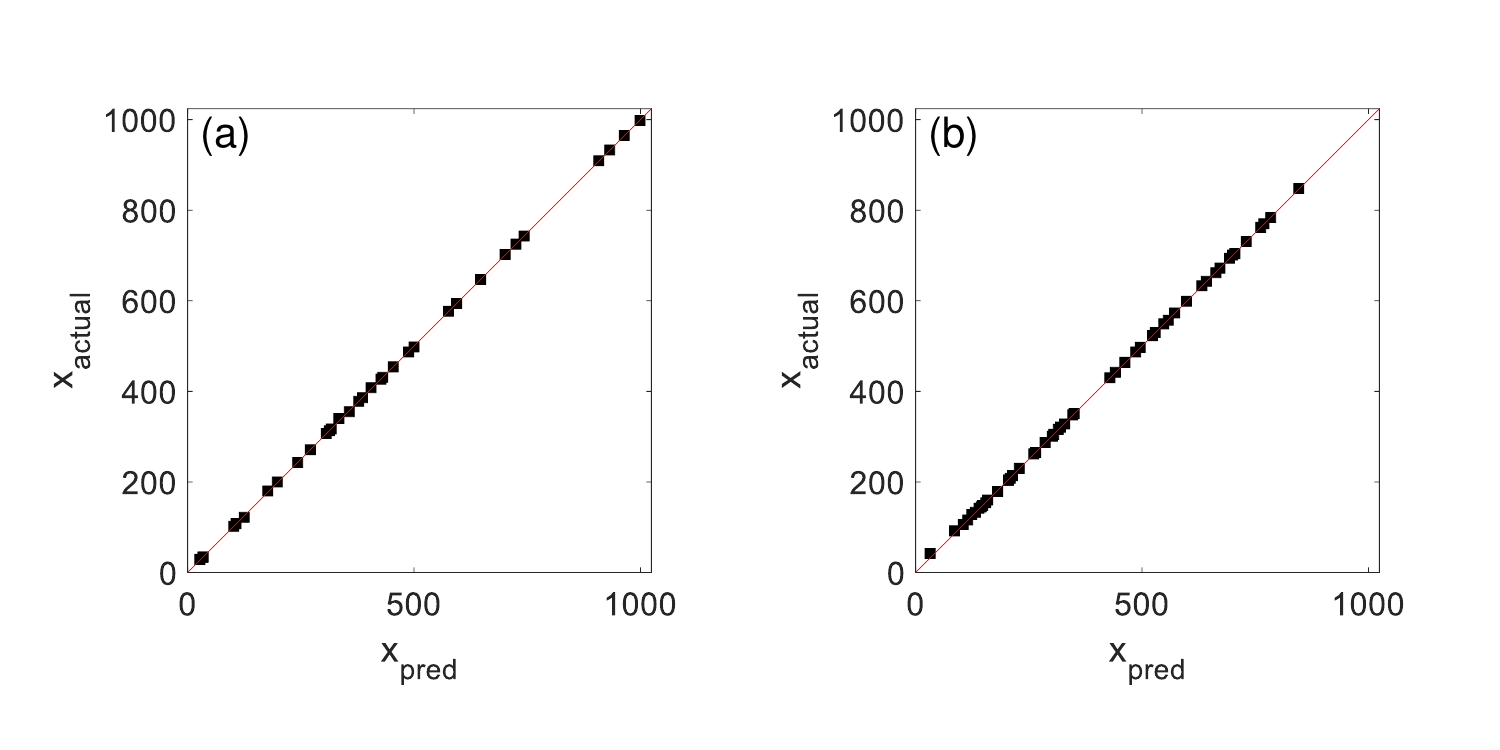}
	\centering\caption{\label{fig:1Dxposition} Correlation between the predicted and computed locations of the localized eigenmodes of 1D multifractal structures generated by (a) the multiplicative cascade method and (b) the MMAR method. The geometrical supports of the eigenmodes are predicted by the landscape function as explained in the text.}
\end{figure}

We then apply the developed approach to the more challenging cases of fractal and multifractal optical potentials that take piecewise constant values equal to $\epsilon_r$ and 1 based on whether a particle is present at a given point of the chain or is missing. The multifractal potentials generated by 1D multiplicative cascade method and by the MMAR method are shown in Figure \ref{fig:1Dcascade}(a,b), respectively. We selected the permittivity of the scattering medium to be $\epsilon_r=10.5$ \cite{DalNegro2021JOSAB}, and the incident wavelength $\lambda$ is chosen to satisfy $\overline{d_1}/\lambda=0.4$ where $\overline{d_1}$ is the average inter-particle separation along the chain. The selected $\overline{d_1}/\lambda$ is typical for the formation of bandgaps in photonic crystal structures \cite{Joannopoulos2008photonic}. 

The landscape functions $u(x)$ calculated by Eq. \ref{eq:1Dlandscape} are plotted in Figure \ref{fig:1Dcascade}(c,d). 
To demonstrate the accuracy of the localized mode predictions from the obtained landscape function, we directly calculated the eigenmodes from Eq. \ref{eq:1Deigenproblem} of both structures and plotted some representative modes in Figure \ref{fig:1Dxposition}(a,b). Particular care has been taken to ensure that the calculated eigenvectors correspond to the eigenvalue of Eq. \ref{eq:1Deigenproblem}, determined by the free-space background medium, which are narrowly distributed around $k_0^2$ (within a $5\%$ dispersion value). 
We observed the coexistence of both localized modes with exponential decay in space and modes with large intensity fluctuations and reduced localization  behavior as well. Moreover, in Figures \ref{fig:1Dxposition}(c,d) we compared the localized mode positions predicted by the peaks of the landscape function $x_\mathrm{pred}$ and the actual positions $x_\mathrm{actual}$ of the modes obtained numerically, selecting a wide range of modes that are spatially distributed along the entire line segment. Our data demonstrate almost perfect correlation between the predicted and the actual positions of the localized modes based on the computed landscape of the Helmholtz operator.

\section{The Helmholtz localization landscape of two-dimensional multifractals}
We now apply the Helmholtz landscape approach to more challenging 2D multifractal systems generated by the  multiplicative cascade method as in reference \cite{Chen2023PRB}. Therefore, we consider the natural generalization of Eq. \ref{eq:1DHelmholtz}: 
\begin{equation}\label{eq:2Deigenproblem}
    H\psi(\mathbf{r})=k_0^2\epsilon_b\psi(\mathbf{r}).
\end{equation}
where the 2D Hamiltonian operator is defined as $H=-\Delta + V^{\prime}(\mathbf{r})$. The associated localization landscape function $u(\mathbf{r})$ is obtained by solving the Dirichlet problem:
\begin{equation}\label{eq:2Dlandscape}
    Hu(\mathbf{r})=1.
\end{equation}
where, similarly to the 1D case, $V^{\prime}(\mathbf{r})\geq0$ everywhere in $\mathbf{r}=(x,y)$.

\begin{figure}[h]
\includegraphics[width=0.9\linewidth]{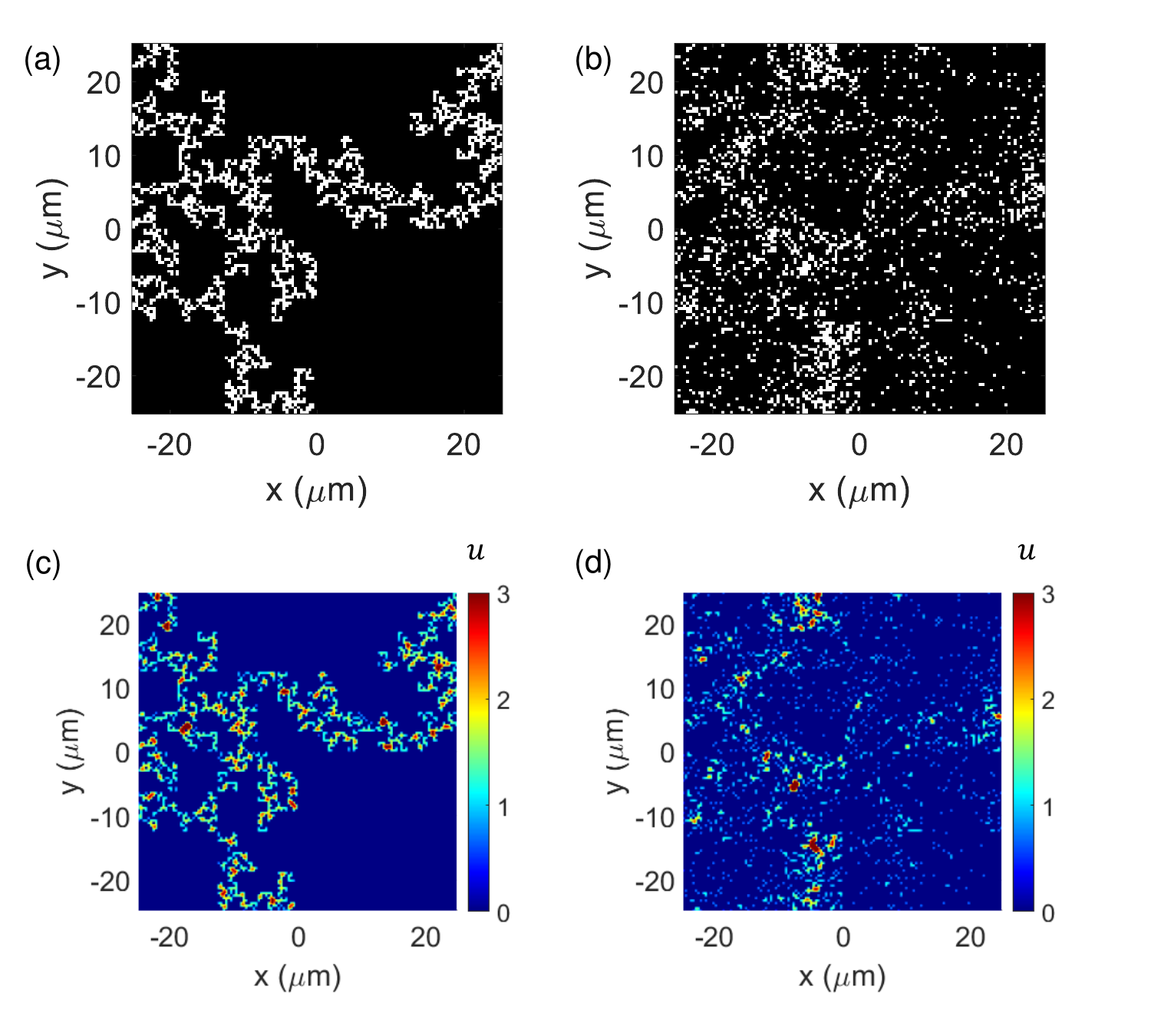}
\centering\caption{\label{fig:2Dlandscape} 2D geometries of the scattering structures corresponding to (a) the monofractal and (b) the multifractal potential. (c,d) Calculated landscape function $u$ of the potentials shown in panels (a) and (b). The spectral parameters used in the calculations are $\overline{d}_1/\lambda=0.55$ and $\overline{d}_1/\lambda=0.62$, respectively.}
\end{figure}

The considered 2D scattering arrays are obtained from multiplicative cascade processes with initial probability vectors that correspond to monofractal (i.e., single-scaling fractals) and multifractal structures. In particular, we considered the case of an initial probability vector $p=[1,1,1,0]$ that produces a monofractal pattern and $p=[1,0.75,0.5,0.25]$ that results is a strongly inhomogeneous multifractal pattern \cite{Chen2023PRB}. The point patterns corresponding to the generated probability random fields and obtained via Monte Carlo rejection are shown in Figures \ref{fig:2Dlandscape}(a,b). Considering the typical dimensions of devices based on nanophotonic membranes, we restrict the multifractal structures within a spatial domain $\Omega$ of $50\times50\,\mu\mathrm{m}^2$, corresponding to approximately 2000 scattering points. The considered potentials are proportional to piecewise permittivity spatial distribution and obtained by dividing $\Omega$ into $128\times128$ unit sub-squares, and we considered a binary potential with value $\epsilon_r$ to each occupied position representing the dielectric scatterers and otherwise with value 1 which represents the air.

The landscape functions are calculated by solving Eq. \ref{eq:2Dlandscape}.  
We wish to setup the Laplacian operator with homogeneous Dirichlet boundary conditions on an open bounded domain $\Omega \in \mathbb{R}^2$ with boundary $\partial\Omega$.
We discretize using finite elements with linear basis functions. Given a bounded, symmetric bilinear form\footnote{in our case, $a(u,v) = \int_\Omega \nabla u \cdot \nabla v$} $a(u, v)$ that is coercive on $H^1_0(\Omega)$, we want to find $u \in H^1_0(\Omega)$ such that $u$ satisfies
\[
a(u, v) = 0, \forall v \in H^1_0(\Omega),
\]
where $H^1_0(\Omega) \subset L_2(\Omega)$, denotes the subspace of functions with square integrable derivatives that vanish on the boundary. This problem is known to have a unique solution $u^*$ \cite{brenner2008mathematical}. For the numerical comparisons in this work we consider domains that are the image of a square under a diffeomorphism, i.e., a smooth mapping from the reference domain $S := [0, 1]^2$ to the physical domain $\Omega$. Quadrilateral finite element meshes and tensorized nodal basis function based on Legende-Gauss-Lobotto (LGL) points are used. We use isoparametric elements to approximate the geometry of $\Omega$, i.e., on each element the geometry diffeomorphism is approximated using the same basis functions as the finite element approximation. The Jacobians for this transformation are computed at every quadrature point, and Gauss quadrature is used to numerically approximate integrals. We restrict our comparisons to uniformly refined conforming
meshes and our implementation, written in Matlab, is publicly available \cite{homg}. It does not support distributed memory parallelism, and is restricted to conforming meshes that can be mapped to a square (in 2D) or a cube (in 3D). While, in practice, matrix assembly for high-order discretizations is discouraged, we use sparse assembled operators in this prototype implementation.
In order to obtain more accurate results, we oversampled the domain $\Omega$ where $2\times2$ elements per constant piece of the optical potential. We need to emphasize that the optical potentials just refers to the coordinates of scatterers and do not take into account the radii of each scatterer. That is to say we treat the scattering medium to be a point distribution, which is not the case for the fabricated nanohole membranes. However, such approximation does not affect the accuracy of the capability of the method to predict eigenmodes because the fabricated nanoholes have very small radii comparing to the multiscale nature of the scatterer separation. The multifractal potential term is added to the discrete Laplacian operator to obtain the discrete Helmholtz operator $H$. Our numerical simulations were performed using the Boston University’s Shared Computing Cluster (SCC) \cite{buscc}.

\begin{figure}[h!]
\includegraphics[width=0.9\linewidth]{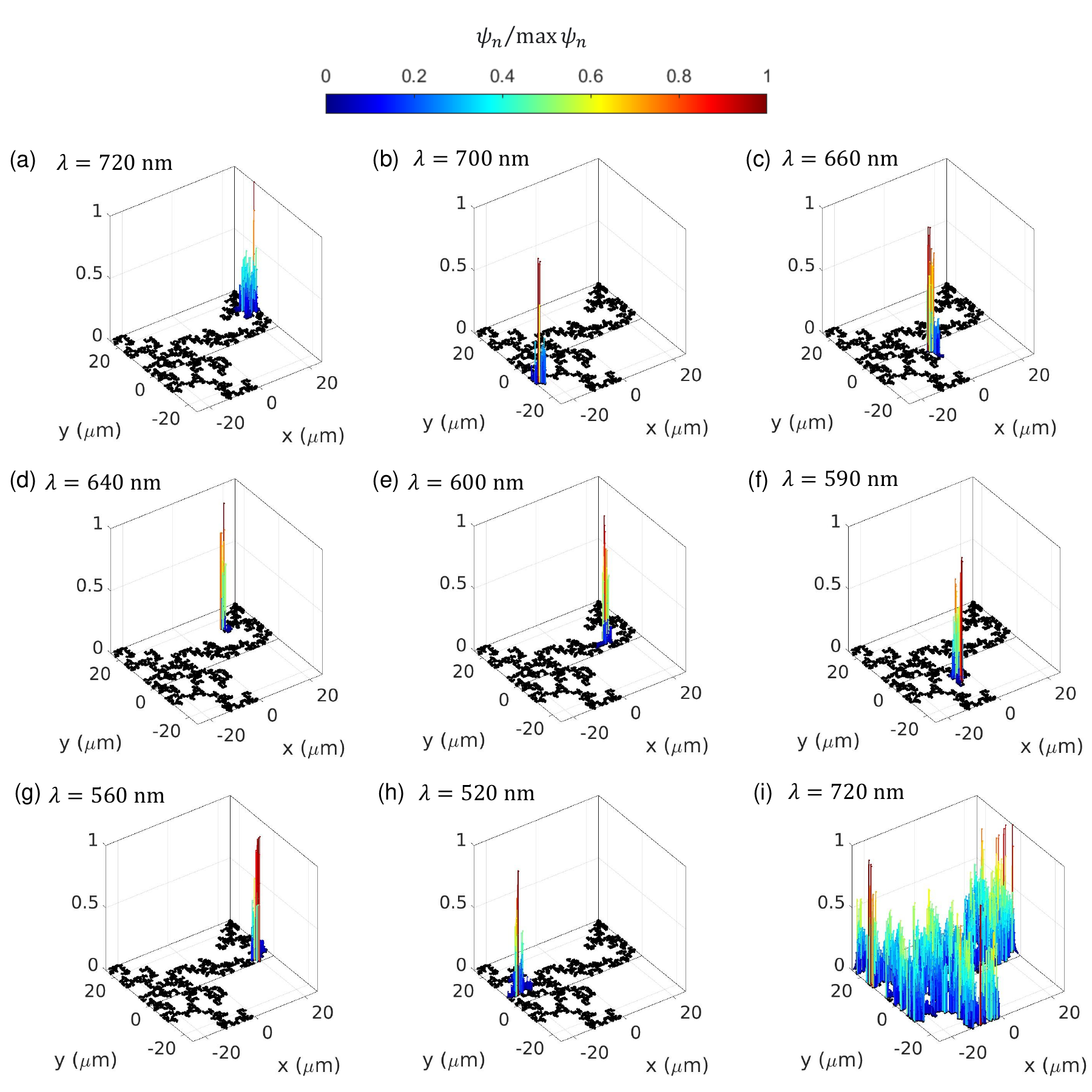}
\centering\caption{\label{fig:2Dmonofractal} (a-h) Eigenmodes whose spectral parameters vary in the range $0.55<\overline{d}_1/\lambda<0.8$. (i) Superposition of eigenmodes corresponding to a narrow range (within a $5\%$ dispersion value) of eigenvalues around $k_{0}^{2}$. The wavelengths $\lambda$ at which the modes are computed are indicated in each panel. }
\end{figure}

In our simulations, we selected the material permittivity to be $\epsilon_r=4$, corresponding to the value for the silicon nitride (SiN) material used in the fabrication of our devices. We chose $\overline{d}_1/\lambda=0.55$ and $\overline{d}_1/\lambda=0.62$ to compute the landscape and the eigenmodes of monofractal and multifractal systems respectively. The selected $\overline{d}_1/\lambda$ lies in the typical range for a photonic crystal structure where bandgaps are likely to form \cite{DalNegro2021JOSAB,Joannopoulos2008photonic} and optical modes with high-quality factors appear at the band-edges  \cite{DalNegro2021JOSAB,trojak2020cavity,trojak2021cavity}. We emphasize that the range of the considered spectral parameter $\overline{d}_1/\lambda$ used in the simulations overlaps with the one of the fabricated structures that will be investigated in the experimental section of this paper. 

\begin{figure}[h]
\includegraphics[width=0.9\linewidth]{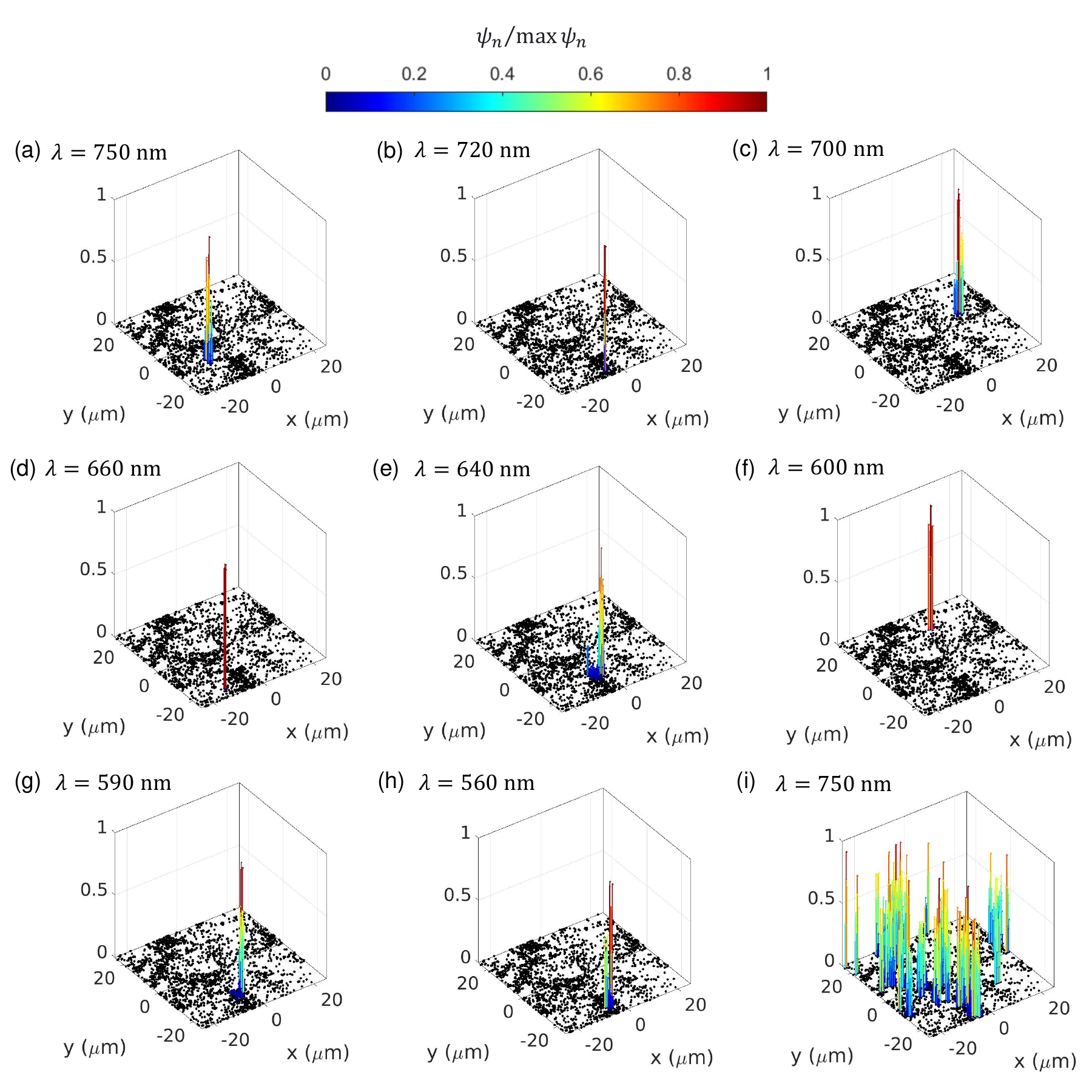}
\centering\caption{\label{fig:2Dmultifractal} (a-i) Eigenmodes whose spectral parameters vary in the range $0.6<\overline{d}_1/\lambda<0.78$. (i) Superposition of eigenmodes corresponding to a narrow range (within a $5\%$ dispersion value) of eigenvalues around $k_{0}^{2}$. The wavelengths $\lambda$ at which the modes are computed are indicated in each panel. }
\end{figure}

The landscape function $u(\mathbf{r})$ of both monofractal and multifractal structures are plotted in Figures \ref{fig:2Dlandscape}(c,d).  One can see that the landscape of the monofractal potential features intense maxima localized within small groups of locally-symmetric clusters of adjacent scattering particles distributed across the structure with spatial distributions that closely resemble its fractal support. These are the regions in which the localization landscape predicts the existence of highly confined resonant modes localized by proximity effects occurring among closely coupled particles. This is the typical behavior observed in small fractal aggregates of dipolar particles \cite{shalaev1999nonlinear}. On the other hand, the landscape of the multifractal potential displays localization regions that are more broadly distributed across the support of the structure. This is consistent with the non-homogeneous multiscale nature of multifractal systems \cite{Chen2023PRB}. In Figures \ref{fig:2Dmonofractal}(a-h) and \ref{fig:2Dmultifractal}(a-h), we display the calculated eigenmodes at the wavelengths specified in each panel. As for the 1D simulations, particular care has been taken to ensure that the eigenvectors are narrowly distributed around $k_0^2$ (within a $5\%$ dispersion value). Moreover, in order to facilitate comparisons with the experimental measurements in which multiple modes are simultaneously excited by an external source, we also plotted in Figures \ref{fig:2Dmonofractal}(i) and \ref{fig:2Dmultifractal}(i) a linear combination of the modes that 
independently contribute to the overall intensity distribution at $\lambda=720$ nm and $\lambda=750$ nm, respectively. As predicted by the localization landscape, we obtain strongly localized modes within small particle clusters for the monofractal structure, which are typical of self-similar arrays where they have been successfully exploited to dramatically increase the cross sections of nonlinear and Raman processes \cite{shalaev1999nonlinear}. However, the situation is quite different in the multifractal structure, where high-intensity localized modes appear to spread across a more extended region of the geometrical support, reflecting its larger degree of spatial non-uniformity compared to the ones of monofractals. This is evident by looking at the mode superposition in Figure \ref{fig:2Dmultifractal}(i) in which the amplitude peaks cover a larger fraction of the structure. In the next sections, we discuss the fabrication and characterization of dielectric membrane structures designed based on the localization landscape theory and we demonstrate experimentally the formation of characteristic fractal modes in qualitative agreement with the landscape predictions.

\section{Fabrication and characterization of multifractal photonic membranes}\label{experimental}
\subsection{Silicon nitride thin film growth and optical characterization}

\begin{figure}[h]
\includegraphics[width=0.8\linewidth]{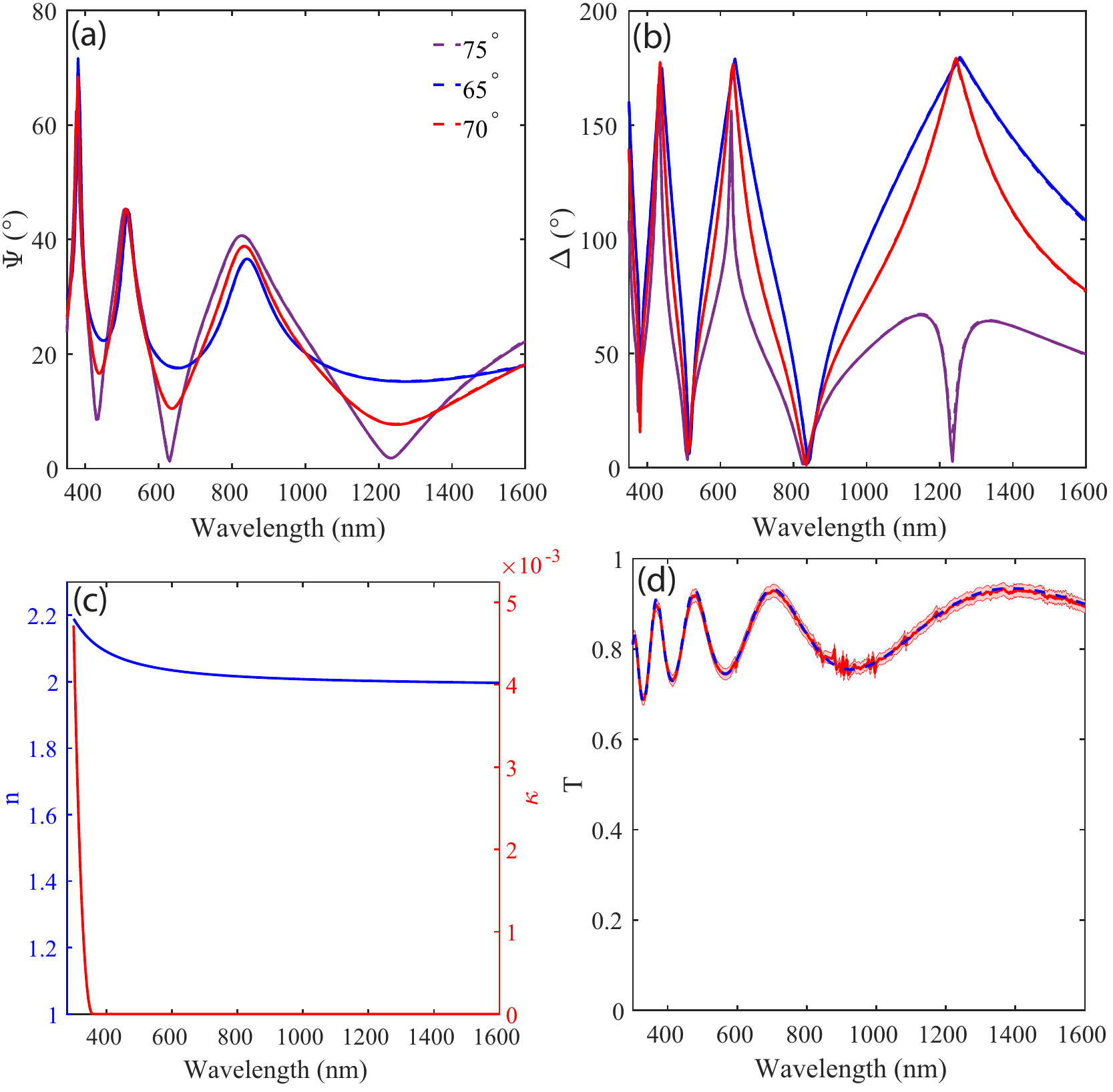}
\centering\caption{\label{fig:Ellipsometry_Sin} Optical properties of fabricated 350nm thick SiN thin films. Ellipsometric parameters (a) $\Psi$ and (b) $\Delta$ as measured (dashed) and fit (solid) at three angles (65$^\circ$,70$^\circ$,75$^\circ$) of the deposited SiN thin films. (c) Refractive index (blue) and extinction coefficients (red) obtained from ellipsometric fitting parameters. (d) Transmission spectra at normal incidence measured (red) and fit (blue-dashed). Error bars of the measurement are represented by the red shading.}
\end{figure}

In this section we discuss the optical characterization of highly transparent silicon nitride thin films grown by reactive magnetron sputtering. Silicon nitride has become a well established material platform for integrated Si photonics where devices such as ultralow-loss waveguides \cite{Bauters:11}, high-Q resonators, and integrated structures for telecommunications, sensing, and metrology have been demonstrated \cite{Xiang:22}. In this work, silicon nitride thin films were grown atop of silicon and fused silica substrates via reactive RF magnetron sputtering using a Denton Discovery 18 sputtering system with a base pressure of $2\times10^{-7}$ Torr and a substrate temperature held at $300^\circ$C. We used a 99.99\% purity 3-in silicon target and sputtered the thin films in a 2:1 argon-nitrogen environment at 2.5 mTorr deposition pressure and 200 W of RF power resulting in a deposition rate of $\approx 5\,\mathrm{nm/min}$. All substrates were washed in a hot piranha solution and plasma ashed in an oxygen environment prior to deposition. We characterize the optical constants of the fabricated SiN thin films from the ultra-violet (UV) to the near infrared (NIR) via broadband variable angle spectroscopic ellipsometry (VASE) and normal incidence transmission measurements. In spectroscopic ellipsometry,  a parameterized oscillator model is introduced to capture the relevant aspects of a materials optical dispersion as a function of wavelength. The oscillator model is then used to interpret the measured data and extract the optical constants of materials via accurate fitting procedures \cite{woollam_1999}. VASE measurements at three angles (65$^\circ$,70$^\circ$,75$^\circ$) were performed on SiN thin films grown atop of silicon substrates along with normal incidence transmission measurements of the same films grown atop of fused silica substrates. Figure \ref{fig:Ellipsometry_Sin} summarizes our work on characterizing the optical properties of the fabricated SiN thin films. Panels (a-b) show the measured and fitted ellipsometric parameters $\Psi$ and $\Delta$ corresponding to the magnitude and phase of the ratio of the complex s- and p-polarized Fresnel reflection coefficients, respectively, at each angle along with each respective fit. The complex Fresnel reflection coefficients are related to the complex refractive index and thickness of the material through parameterized oscillator models. In particular, we employ a single parameterized Tauc-Lorentz (TL) oscillator model which has been used to extract the optical constants of amorphous materials in excellent agreement with experimental measurements \cite{shubitidze_2023}. We find that this modeling method gives nearly ideal agreement with our measurements over the entire UV-VIS-NIR region investigated with a mean square error (MSE) of less than 10. In Figure \ref{fig:Ellipsometry_Sin}(c) we report the extracted optical constants of the fabricated 350 nm thick SiN thin film as a function of wavelength. Furthermore, we performed normal incidence transmission measurements on the SiN thin films grown atop of fused silica substrates and independently test our TL model by plotting the expected transmission. We report our measurements and fit in fig \ref{fig:Ellipsometry_Sin}(d) finding the TL model is in excellent agreement with measurement over the entire wavelength range investigated. The transmission spectra, along with the extracted optical constants of the fabricated material, demonstrates exceptionally high transparency with an almost constant refractive index of $2$. Consistent with literature, we attribute these optical properties to the high RF power and substrate temperature during deposition\cite{hegedus2021}. In the next section we discuss the fabrication of nanohole perforated membranes with multifractal geometries.

\subsection{Multifractal photonic membrane fabrication}
We fabricate multifractal nanohole arrays within the 350 nm thick SiN thin film via electron beam lithography (EBL) and anisotropic reactive ion etching (RIE). 
\begin{figure}[h]
\centering\includegraphics[width=0.83\linewidth]{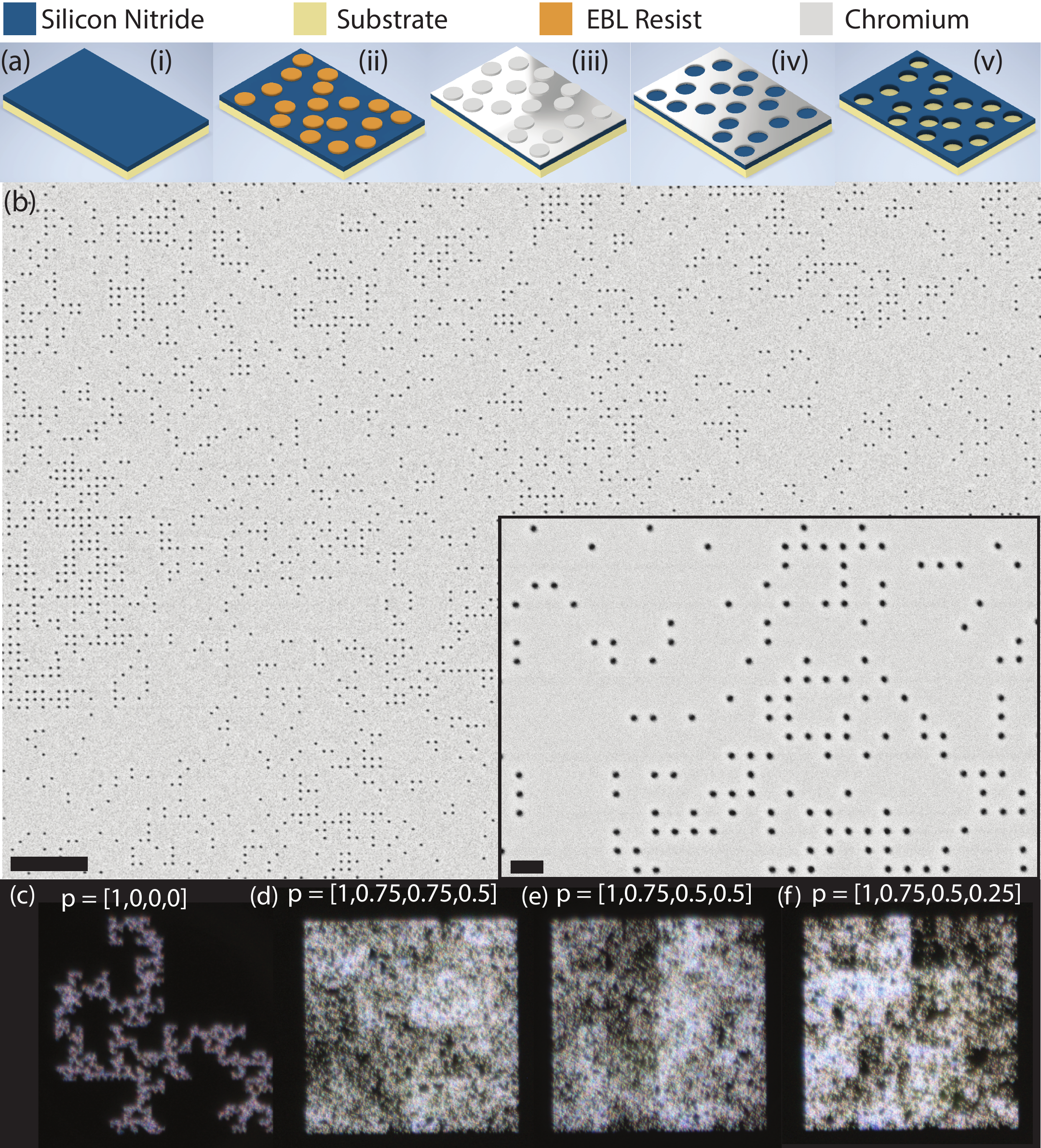}
\caption{\label{fig:Fab_DF_SEM} (a) Fabrication process flow via EBL and anisotropic RIE. (b) Representative SEM image of the fabricated multifractal pattern. The inset is a close up view of the same sample. The scale bars in the SEM image and inset are 4 $\mu m$ and 600 nm respectively. (c-f) Dark field scattering images of monofractal and multifractal structures. Each array consists of $\approx$15,000 points written over a 100 $\mu m^2$ area. The initial probability vector used to generate each pattern is reported above each image. }
\end{figure}
A sketch of the process flow is shown in Figure \ref{fig:Fab_DF_SEM}(a). First a negative tone resist is spun atop of the grown material to achieve a thickness of $\approx$100 nm followed by soft bake.
A thin, conductive coating is then spun atop of the resist to dissipate charging during the EBL process. Nano-hole patterns are then written onto the photoresist via EBL using a 30kV source (Zeiss Supra40VP). To develop the patterns, the sample is rinsed with DI water to remove the protective conductive coating and then submerged in a 4:1 ratio of developer and DI water for 55-60 seconds and blow dried with N$_2$ gas. A brief plasma ashing step in a pure oxygen environment is performed in order to remove residual resist left over from the development stage. This step is crucial in achieving reliable liftoff. A 20 nm chromium mask is then evaporated onto the sample (Step 3) and submerged in acetone for lift-off. A dry etch of the exposed nano-holes is performed using anisotropic reactive ion etching and finally a wet chromium etch is performed to reveal the final nano-hole structure. Representative scanning electron microscope (SEM) images of a multifractal sample are shown in Figure \ref{fig:Fab_DF_SEM}(b) along with dark-field optical images of all multifractal structures fabricated  are shown in Figures \ref{fig:Fab_DF_SEM}(c-f). All patterns were written onto a square 100 $\mu\mathrm{m}^2$ area with each hole having the desired radius of 100 nm as confirmed by SEM imaging. In order to facilitate the comparison of the measured and simulated optical modes across the investigated spectral range, both monofractal and multifractal membranes were fabricated with $\overline{d}_1/\lambda=0.62$ at the operation wavelength $\lambda=700\,\mathrm{nm}$.

\subsection{Experimental characterization of multifractal resonances}
In order to demonstrate the excitation of polarization-resolved multifractal modes of the fabricated samples, we utilize a leaky mode imaging setup \cite{Mondal2019}. Our experimental setup uses a broadband COMPACT K super-continuum laser source filtered by a monochromator (Oriel Cornerstone 260) to achieve a spectral linewidth of 2 nm. The filtered light is then TE polarized using a linear polarizer, collimated and focused onto the edge of the sample by an objective lens.
\begin{figure}[h]
\includegraphics[width=0.85\linewidth]{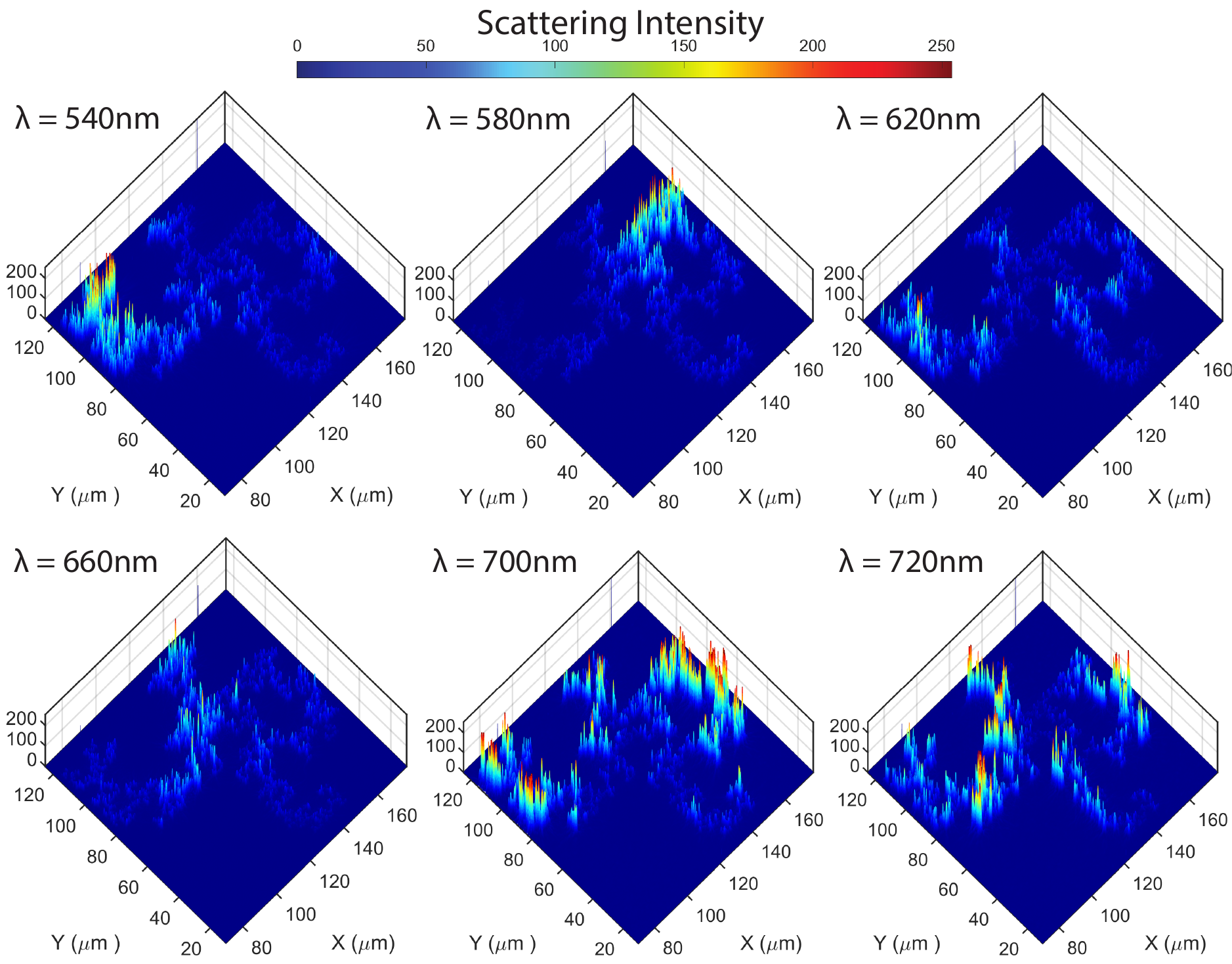}
\centering\caption{\label{fig:exp_monofractal} Representative leaky mode measurements of a monofractal nanohole array fabricated in SiN at various excitation wavelengths. The excitation wavelength is indicated by $\lambda$ in each panel. The corresponding spectral parameter varies in the range  $0.6<\overline{d}_1/\lambda<0.8$.}
\end{figure}
The out of plane scattered radiation is collected by a high NA objective (50x) and an image of the modes is focused onto a camera (Thorlabs CS126MU) using a tube lens. 
We first measured the leaky modes of the fabricated monofractal structure as a function of wavelength from 450 nm-750 nm. In Figure \ref{fig:exp_monofractal} we show representative images of the spatial distributions of the optical modes of the monofractal structure excited using TE-polarized light at different wavelengths. We observe clearly distinct spatial distributions of the modes when excited at different wavelengths. Moreover, all mode patterns from these data feature spatial localization within small clusters of particles with highly fluctuating intensity profiles that are qualitatively similar to the numerical predictions from the landscape theory. This characteristic modal clustering behavior around smaller regions of space of the monofractal geometry is also manifested by the calculated modes in Figure \ref{fig:2Dmonofractal} and is expected for self-similar structures in which scattering resonances are driven by coupling effects within locally-symmetric clusters distributed at multiple length scales \cite{shalaev1999nonlinear,Sgrignuoli2020}. 

As a comparison, we also investigated the optical resonances of the multifractal structure generated  with the probability vector $p=[1,0.75,0.5,0.25]$ and a spectral parameter identified using the 
landscape theory, enabling a direct experimental comparison of the leaky-mode structures of fractals and multifractal systems at minimal computational cost. 
\begin{figure}[h]
\includegraphics[width=0.85\linewidth]{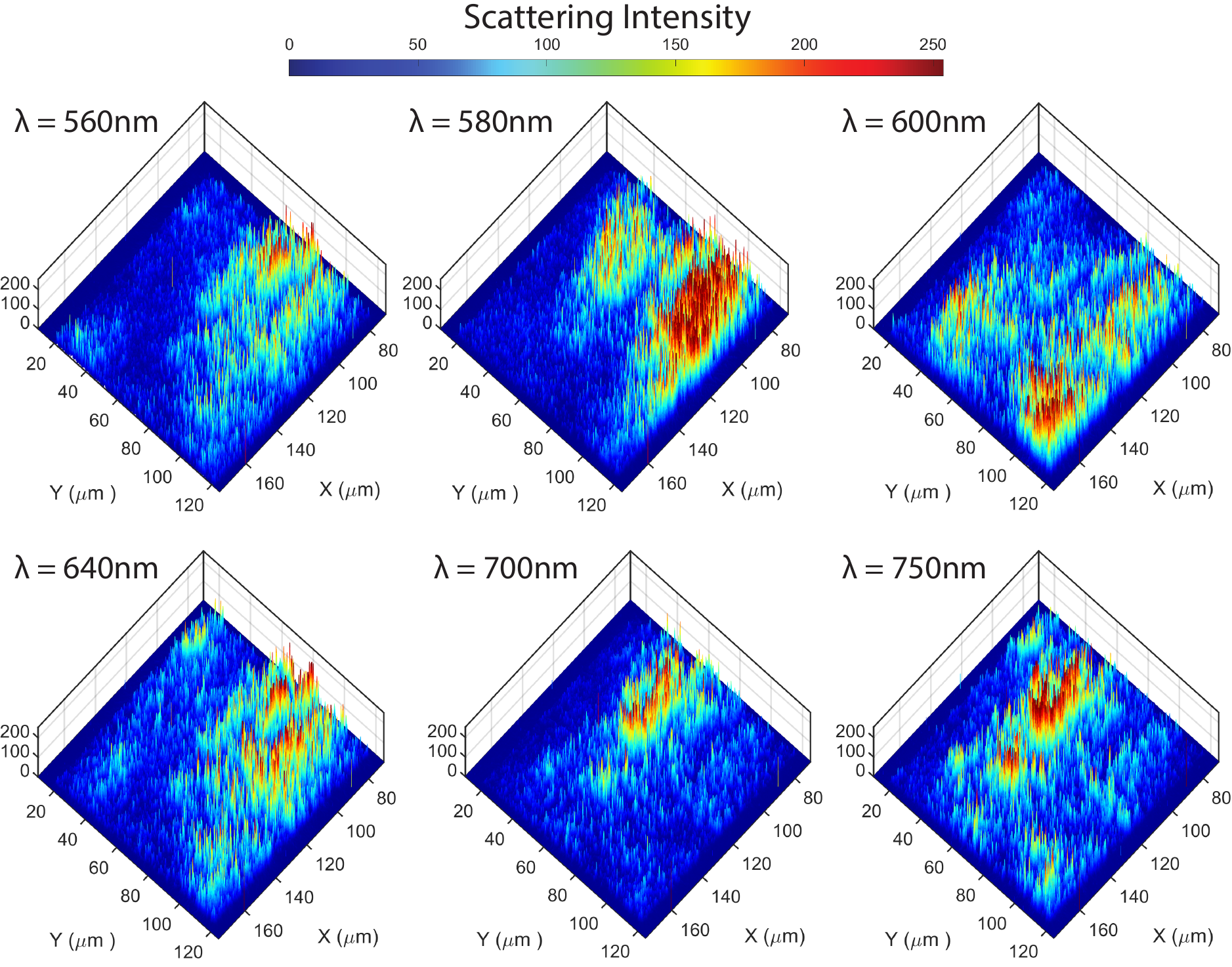}
\centering\caption{\label{fig:exp_multifractal} Representative leaky mode measurements, at various wavelengths, of a multifractal nanohole array fabricated in SiN generated with an initial probability vector of p = [1, 0.75,0.5,0.25]. The excitation wavelength is indicated by $\lambda$ in each panel. The corresponding spectral parameter varies in the range  $0.6<\overline{d}_1/\lambda<0.78$.}
\end{figure}
As shown in in Figure \ref{fig:exp_multifractal}, the observed mode patterns of the multifractal structure display highly fluctuating spatial profiles distributed across the entire geometrical support, in qualitative agreement with the simulation results shown in Figure \ref{fig:2Dmultifractal}.  Moreover, due to the spatially non-homogeneous nature of the investigated multifractal structure, the measured modes display no evident trend in the degree of their spatial localization when varying the wavelength, which is indicative of a broader spectrum of localized resonances compared to monofractals \cite{Chen2023PRB}. We emphasize that, since the Helmholtz landscape method introduced here relies on scalar fields and point-like particles, it can only offer basic insights into the general nature of the localized modes. Specifically, these insights relate to the ability to pinpoint exactly the localization regions of the modes and predict global properties such as the localization/delocalization behavior of the resonant states. Therefore, it is unreasonable to expect a quantitative agreement with the experimental data at this stage. In particular, the model developed here does not capture the effects of the finite-size of the scattering particles, which may introduce local (small-scale) modifications to the overall mode structures compared to the ones measured on the fabricated samples. However, our work shows that the landscape approach is still valuable in providing predictive insights at low computational cost in large-scale scattering structures with complex potentials that are often beyond the reach of fully numerical, grid-based techniques such as the finite element method. Thus, we consider the proposed approach as a viable tool for the efficient first-order design of complex photonics structures.
Future work will focus on the quantitative analysis of the spectral measures of the investigated multifractal optical systems, such as their integrated density of states \cite{Arnold2019SIAM}, and the statistical distribution of intensity maxima of eigenvectors that unveil long-range multifractal correlations beyond the traditional Anderson model  \cite{mirlin1996transition,fyodorov1995statistical,falcao2022wave}.

\section{Conclusion}
In this paper, we proposed a method to efficiently design extended photonic structures with multifractal geometries and investigate the fundamental properties of the fractal eigenmodes of 1D and 2D Helmholtz operators with multifractal scattering potentials. In particular, we explored two canonical multifractal systems generated from the multiplicative cascade and the MMAR methods. Without solving the associated eigenproblems, we calculated the landscape functions and accurately predicted the locations of the supported eigenmodes. Finally, based on the information obtained from the localization landscapes, we designed and fabricated multifractal photonic membranes in highly transparent SiN materials and directly imaged, using leaky-mode spectroscopy, their optical modes across the visible spectral range. The general predictions from the localization landscape of the Helmholtz operator were found to be in good qualitative agreement with the experimental data, establishing the localization landscape theory as a viable tool for the rapid exploration and benchmarking of scattering resonances in complex photonic structures with tailored multifractal disorder for nanophotonics and metamaterials applications.
Finally, our results unveil the distinctive localization behavior of the scattering resonances supported by extended fractal and multifractal photonic membranes. 

\begin{backmatter}
\bmsection{Acknowledgments}
L.D.N. acknowledges support from the National Science Foundation (ECCS-2015700, ECCS-2110204) and insightful discussions with Marcel Filoche and Sergey Skipetrov.

\section*{Disclosures}
 The authors declare no conflicts of interest.
 
\section*{Data availability} 
Data will be available upon reasonable request. 

\end{backmatter}


\bibliography{sample}






\end{document}